\documentclass[12pt]{article}
\usepackage[dvips]{graphicx}
\newcommand{\beq}{\begin{equation}}
\newcommand{\eeq}{\end{equation}}
\newcommand{\beqa}{\begin{eqnarray}}
\newcommand{\eeqa}{\end{eqnarray}}
\newcommand{\IP}{I\!\!P}
\begin{document}  
\title{RHIC and diffraction in $pp$ 
Spin-flip\thanks{Work supported in part by the MURST of Italy}.}
\author{A.F. Martini\thanks{Fellow of FAPESP, SP, Brazil}
\thanks{e-mail: martini@to.infn.it,martini@ifi.unicamp.br} 
 and E. Predazzi\thanks{e-mail: predazzi@to.infn.it} \\
Depto.  Fisica Teorica, Univ. of Torino - Italy \\
and \\ 
INFN - Sezione di Torino - Italy
}
\date{}
\maketitle
\abstract{ We argue that diffraction (Pomeron) contribution is present 
in the $pp$ spin-flip amplitude. RHIC polarization data 
will be able to prove (or disprove) this conjecture.}
\section{Introduction}
\label{sec:intro}

Following the hints given in very old pioneer papers
\cite{predazzi67,hinotani79}, in a recent work \cite{martini00} we 
discussed the possible presence of a
diffractive-like (Pomeron) contribution in the ``{\it reduced}'' 
spin-flip amplitude, that is, the spin-flip amplitude when the kinematical 
zero is removed. The procedure used was to remove this zero 
factorizing a $\sin\theta$ factor (eq. (6) of \cite{martini00}) 
according to the suggestion of \cite{predazzi67,hinotani79}.

The result obtained in \cite{martini00} predicted a very small
polarization $P$ at the Relativistic Heavy Ion Collider (RHIC) 
energies but, upon reconsideration of the problem, it is necessary to
verify how much may this extrapolation is influenced by the 
use of the factor $\sin\theta$\footnote{We are grateful to the referee 
and to Prof. O.V. Selyugin for stressing this point to us.}, 
since this is not the only way 
to remove the kinematical zero. This could, alternatively, be 
done using a $\sqrt{-t}$ factor 
\cite{buttimore99,akchurin99,goloskokov91} which entails a $\sqrt{s}$ 
factor as compared with the $\sin\theta$ option leading, possibly, 
to a significatively larger $P$ at high energies.

Using $\sqrt{-t}$ appears more in line with the Regge behavior 
prescription but only the data can, ultimately, clarify the issue. 
For this, we revise here our conclusions reached in \cite{martini00} 
using $\sqrt{-t}$ to show how RHIC should be able to settle the point.

To investigate $pp$ scattering, it is necessary to specify 
five independent helicity amplitudes in terms of which the polarization 
$P$ is given by
\vskip 0.3cm
\beq
P=2{{\rm Im}[(\phi_1+\phi_2+\phi_3-\phi_4)\phi_5^{\ast}]\over 
[|\phi_1|^2+|\phi_2|^2+|\phi_3|^2+|\phi_4|^2+4|\phi_5|^2]},
\eeq 
where $\phi_1,\phi_3$ are spin-non-flip amplitudes, $\phi_2,\phi_4$ 
are double spin-flip amplitudes and $\phi_5$ is the single spin-flip 
amplitude. 

Following \cite{martini00}, the amplitudes $\phi_2$ and $\phi_4$ 
will be neglected and we write 
\beq
\phi_1\sim g(s,t),\quad \phi_5=h(s,t)
\label{effspamp}
\eeq
assuming $\phi_1=\phi_3$. $g(s,t)$ and $h(s,t)$ will be considered 
{\it effective} spin-non-flip and spin-flip amplitudes, respectively. 
These definitions are not exactly those used in \cite{martini00} but 
this is immaterial since $\phi_1,\;\phi_3$ and $\phi_5$ were not 
explicitly given there.

Using eq. (\ref{effspamp}), $P$ can now be rewritten as
\beq
P=2{{\rm Im}[g(s,t)h^{\ast}(s,t)]\over |g(s,t)|^2+2|h(s,t)|^2} .
\eeq
We use the same set of data used in \cite{martini00} for $pp$ polarization 
data \cite{kline80} and the parametrization for the $pp$ spin-non-flip
amplitude is again taken from reference \cite{desgrolard00}. Removing the
kinematical zero with $\sqrt{-t}$ (see below) we reach the following
conclusions:

\begin{description}
\item{a)} As in \cite{martini00}, we find that the presence of a 
diffractive-like behavior in the (reduced) spin-flip amplitude 
leads to good fit of the data but, differently from the 
$\sin\theta$ case, the reduced spin-flip amplitude is now only about 
10\% of the imaginary part of the spin-non-flip amplitude.
\item{b)} The data fitting with the same energy dependence in $g(s,t)$ and
$h(s,t)$ seems the best choice; the zero of the polarization moves 
towards zero with energy as noticed earlier \cite{martini00}.
\item{c)} The magnitude of the polarization decreases as the 
energy increases but the extrapolation to 500 GeV 
predicts a non-negligible contribution if the same Pomeron 
trajectory for both spin-flip and spin-non-flip amplitudes is used;
\item{d)} if the intercepts of the trajectories of the Pomeron in 
$g(s,t)$ and $h(s,t)$ are not the same, the extrapolation to 500 GeV 
becomes quite unrealistic (and just as small as it was obtained with the 
factor $\sin\theta$). 
\end{description}

Our conclusion is that RHIC will really be able to give a clear cut 
answer to the long standing question: is diffraction (the Pomeron) 
contributing to spin-flip in $pp$?

\section{Definition of the amplitudes}
\label{sec:defin}

The {\it effective} $pp$ spin-non-flip amplitude will be written as
\beq
g(s,t)=a^{nf}(s,t)=a_{+}(s,t)-a_{-}(s,t)
\eeq
with
\beqa
a_{+}(s,t)&=&a_{\IP}(s,t)+a_f(s,t),\nonumber\\
a_{-}(s,t)&=&a_{O}(s,t)+a_{\omega}(s,t)
\eeqa
where $a_{\IP}(s,t)$ and $a_{O}(s,t)$ are the $\IP$ (Pomeron) and Odderon
amplitudes respectively and $a_f(s,t)$ [$a_{\omega}(s,t)$] are the 
even [odd] secondary Reggeons\footnote{Actually, $a_f$ embodies 
both $f$ and $\rho$ contributions (and $a_{\omega}$ both $\omega$ and 
$a_2$).}. These different amplitudes are taken directly from Ref. 
\cite{desgrolard00} and their explicit forms are given in Appendix 
\ref{appe:spin} together with the values of their parameters.

In the effective spin-flip amplitude, $h(s,t)$, we neglect the 
contribution of secondary Reggeons and, {\it mutatis mutandis}, 
we follow \cite{martini00} to write 
\beqa
h(s,t)&=&a^{sf}(s,t) = (i\gamma_1+\delta_1){\sqrt{-t}\over m_p}\;
\tilde{s}^{\alpha^{sf}(t)}e^{\beta_1 t}\Theta(0.5-|t|)   
\nonumber\\
&+& (i\gamma_2+\delta_2){\sqrt{-t}\over m_p}\;
\tilde{s}^{\alpha^{sf}(t)}e^{\beta_2 t}\Theta(|t|-0.5),
\label{spinampl2}
\eeqa
where the mass of the proton, $m_p$, is used to make the parameters 
dimensionless. In (\ref{spinampl2}) 
$\tilde{s}={s\over s_0}e^{-i\pi/2}$, $\Theta$ is the step function 
and we assume $s_0=1\;{\rm GeV^2}$ as in \cite{desgrolard00}; the 
superscript $sf$ (for ``spin-flip'') will allow us to check if 
the $\IP$ trajectory can be different for spin-flip and 
spin-non-flip. In the parametrization (\ref{spinampl2}) the Pomeron 
contribution to the spin-flip amplitude is allowed a complex phase 
and this, we believe, can be justified by the CP-even contribution 
induced by the exchange of three(or more)-gluon ladders.

To start with, we take the spin-flip Pomeron trajectory 
$\alpha^{sf}(t)$ to be exactly the same as derived for 
the spin-non-flip amplitude 
\beq
\alpha^{sf}(t)=\alpha_{\IP}(t)=\alpha_{\IP}(0)+\alpha_{\IP}'t
\label{alfsf}
\eeq
where $\alpha_{\IP}(0)$ and $\alpha_{\IP}'$ are found in Appendix A. The
data at $\sqrt{s}=13.8$, 16.8 and 23.8 GeV (a total of 64 points) are used in
the fit and the values of the parameters, together with the $\chi^2$ are 
listed in Table \ref{tab6param}.

\begin{table}[hbt!]
\centerline{
\begin{tabular}{|c|c|c|c|}
\hline
$\gamma_1$ & $1.35\times 10^{-1}$ & $\gamma_2$ & $2.55\times 10^{-2}$ \\
$\delta_1$ & $2.64\times 10^{-1}$ & $\delta_2$ & $5.38\times 10^{-2}$ \\
$\beta_1\;({\rm GeV}^{-2})$ & 4.74 & $\beta_2   
\;({\rm GeV}^{-2})$ & 2.29 \\ \hline
\multicolumn{4}{|c|}{$\chi^2/d.f.=1.1$}\\
\hline
\end{tabular}
}
\caption{Values of the parameters obtained from fitting
polarization data at $\protect\sqrt{s}=13.8$, 16.8 and 23.8 GeV with 
eqs. (\protect\ref{spinampl2}) and (\protect\ref{alfsf}).
}
\label{tab6param}
\end{table}

\begin{figure}[hbt!]
\begin{center}
\centerline{
\includegraphics[width=8cm,height=7.5cm]{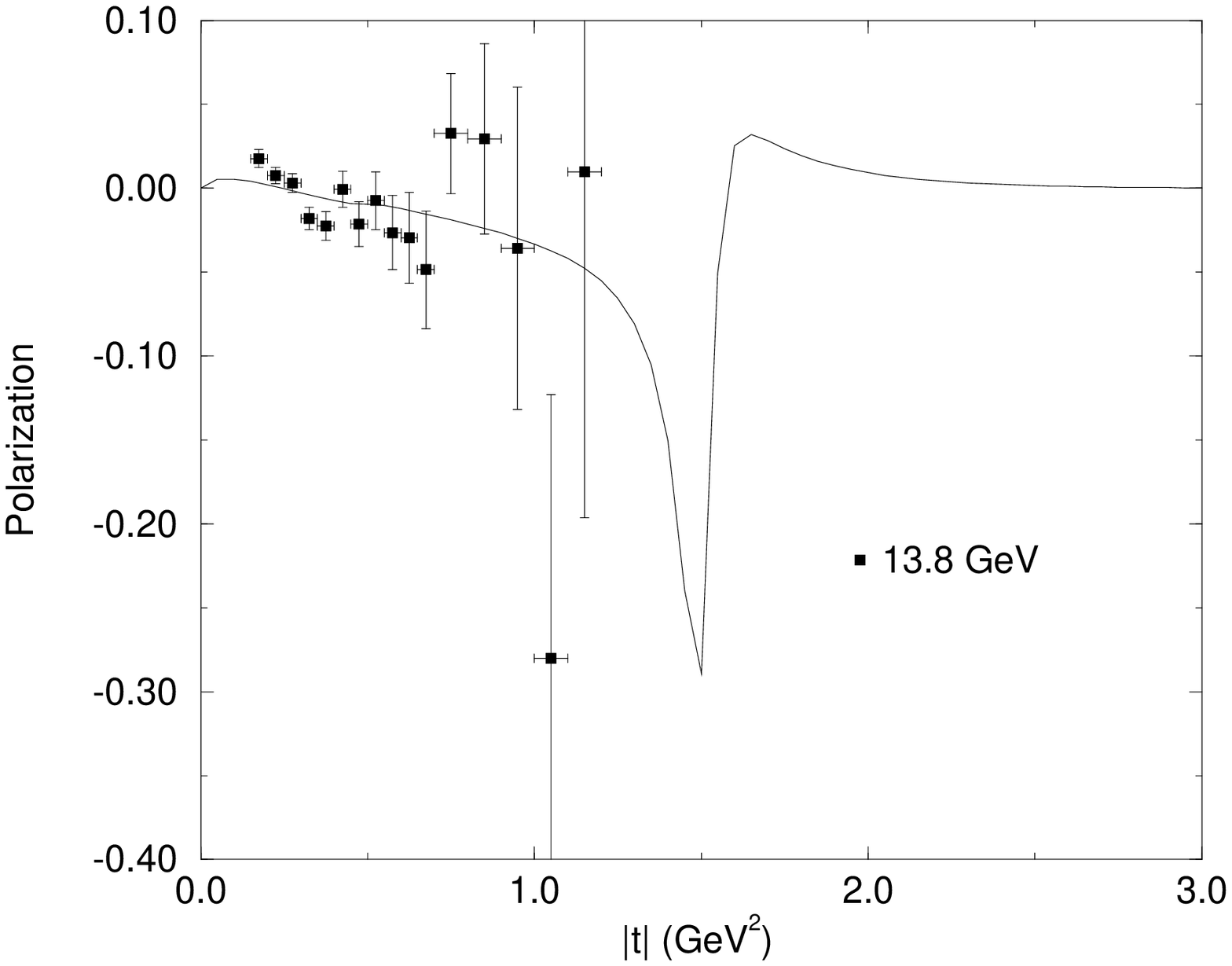}
\includegraphics[width=8cm,height=7.5cm]{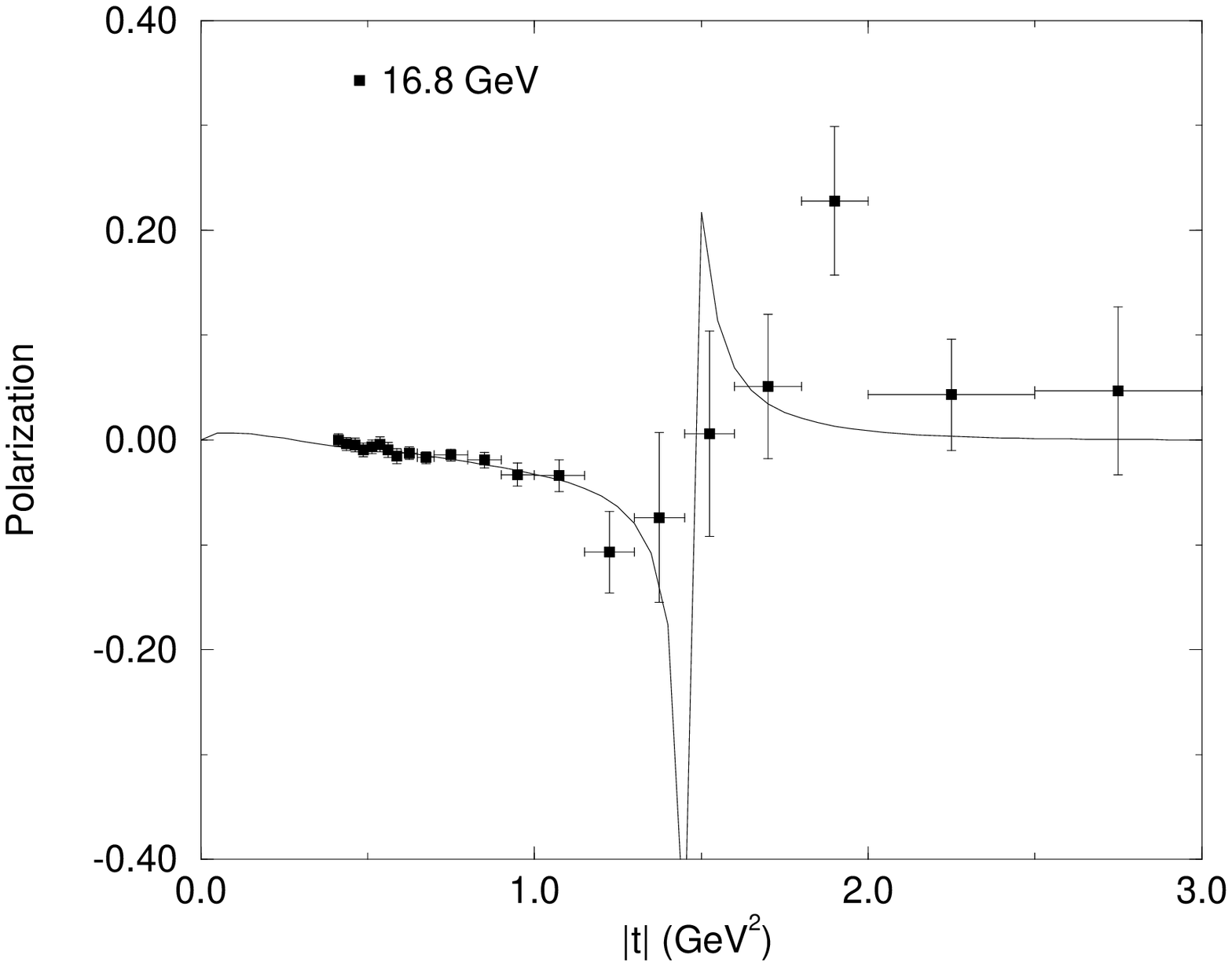}
}
\centerline{
\includegraphics[width=8cm,height=7.5cm]{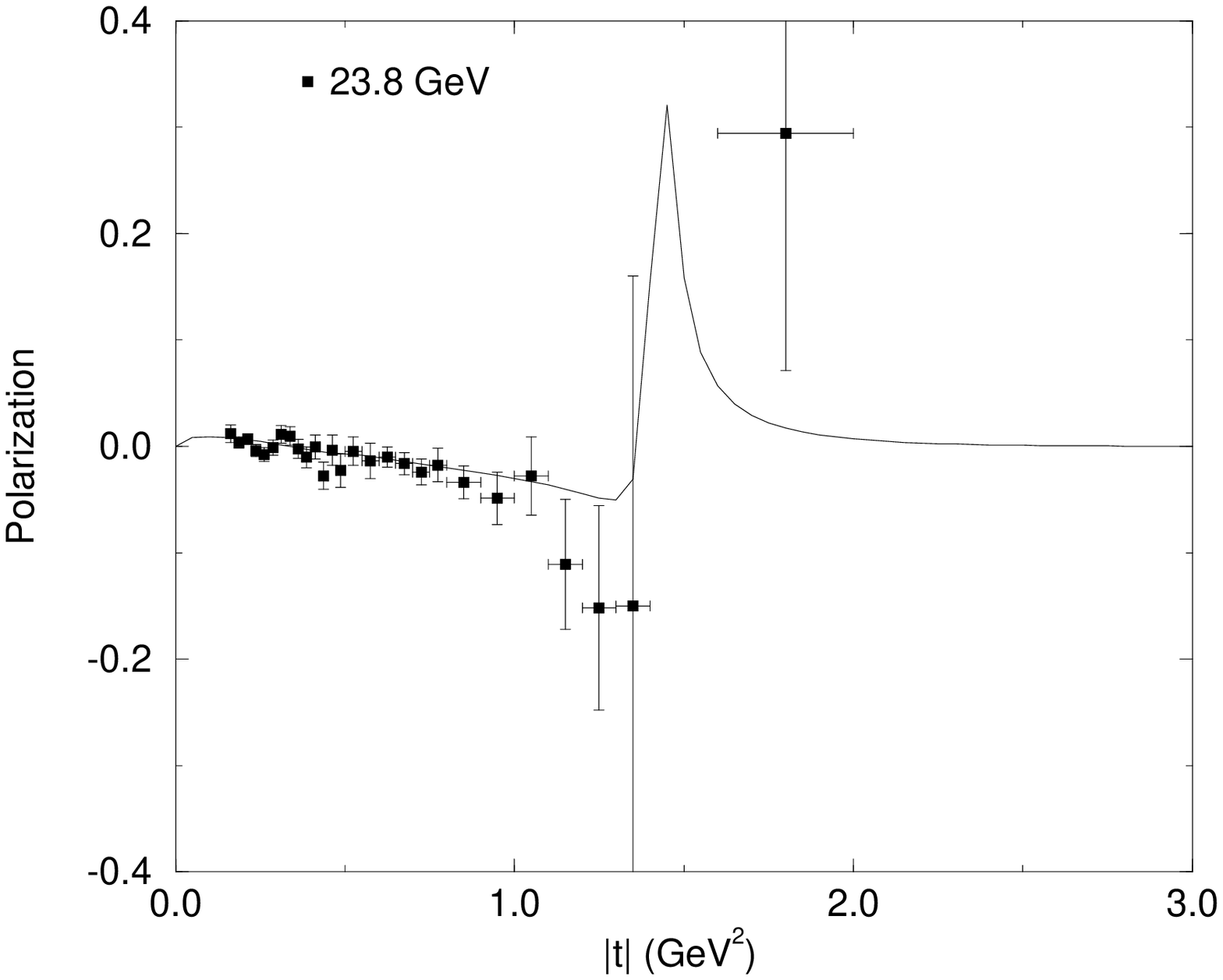}}
\caption{Results from fitting polarization data 
at various energies: 13.8, 16.8 and 23.8 GeV (see Table
\protect\ref{tab6param}).
}
\label{figpol6param}
\end{center}
\end{figure}

\begin{figure}[hbt!]
\begin{center} 
\centerline{
\includegraphics[width=8cm,height=7.5cm]{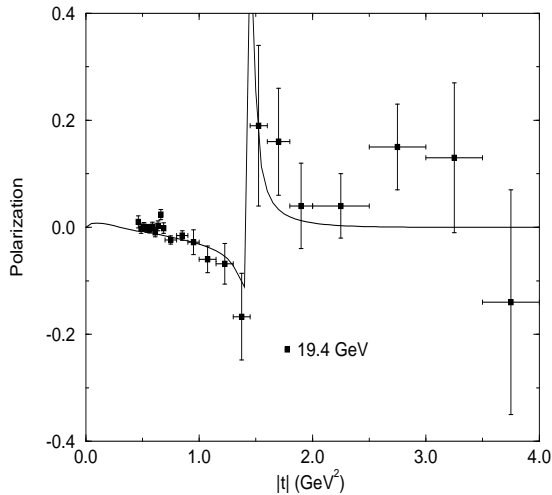}}
\caption{ The prediction for the polarization at 
19.4 GeV (not used in the fit) compared with the experimental data at 
that energy.
}
\label{figpol19pt4}
\end{center}
\end{figure}

\begin{figure}[hbt!]
\begin{center}
\centerline{
\includegraphics[width=8cm,height=7.5cm]{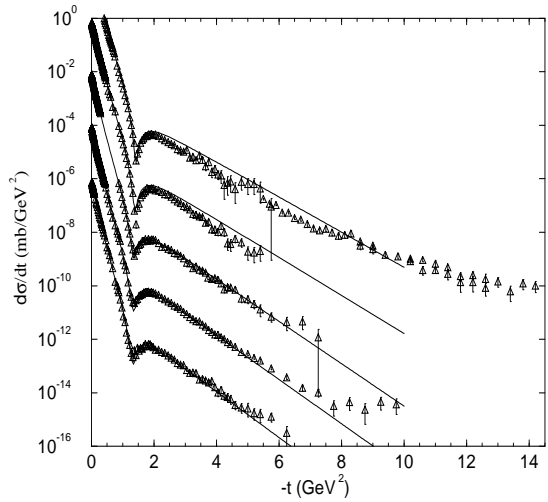}
}
\caption{The differential cross section obtained in this
work taking into account the spin-flip amplitude. The highest set of data
correspond to 23.5 and 27.4 GeV grouped together. The other sets
(multiplied by powers of $10^{-2}$) are 30.5, 44.6, 52.8 and 62 GeV.
}
\label{figdsdt}
\end{center}
\end{figure}

In Fig. \ref{figpol6param} we show the polarization data together 
with our reconstruction. As a check of the validity of our solution, 
Fig. \ref{figpol19pt4} shows how it accounts for the data at 
$\sqrt{s}=19.4$ Gev (not used in the fit). In Fig. \ref{figdsdt} 
we show $d\sigma/dt$ at various energies.

\newpage

\begin{figure}[hbt!]
\begin{center}
\centerline{
\includegraphics[width=8cm,height=7.5cm]{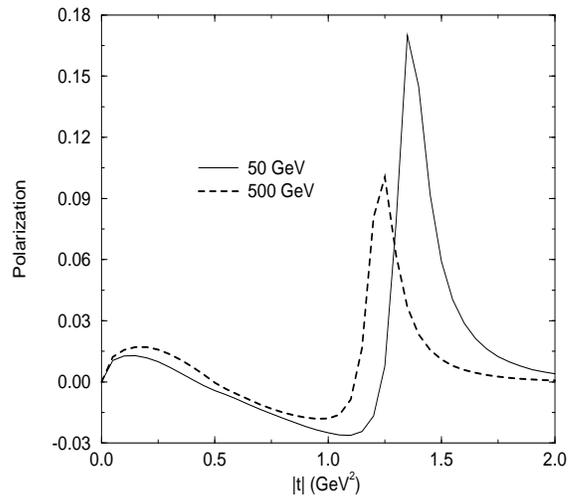}
}
\caption{The polarization predictions for 50 and 500 GeV
with the parameters of Table \protect\ref{tab6param}.
}
\label{figpol50500} 
\end{center}  
\end{figure}

Some considerations are in order:
\begin{description}
\item{a)} the $\IP$ contribution to the spin-flip amplitude is
considerably smaller than to $g(s,t)$ (about 10\%, roughly)
contrary to the $\sin\theta$ case \cite{martini00}; 
\item{b)} the (small $|t|$) slope of the spin-flip amplitude
$\beta_1=4.74\;{\rm GeV}^{-2}$ is somewhat smaller than the  
one determined previously \cite{hinotani79,martini00} 
but this is not unexpected due to the changes made 
in eq. (\ref{spinampl2});
\item{c)} the extrapolation of our solution to 50 and 500 GeV 
predicts the polarization shown in Fig. \ref{figpol50500}. The curve 
at 500 GeV is smaller than that at 50 GeV but considerably larger 
than the prediction with $\sin\theta$ \cite{martini00}; most 
important, this polarization could be measured at RHIC;
\item{d)} our result ({\it i.e.}, $h(s,t)$)
cannot be extended to $|t|$ values much higher than few ${\rm GeV}^2$
because the spin-non-flip amplitude utilized is valid at the Born level
\cite{desgrolard00} and its description in the region after the
dip ($|t|> 1.5\;{\rm GeV}^2$) is not very good. To extend our
considerations to higher $|t|$, it would be 
necessary to adopt the more sophisticated eikonalized version. Anyway,
the $t$-region of interest for RHIC is up to $1.5\;{\rm GeV}^2$
\cite{guryn00} so we can confine our analysis to the not too high
$|t|$-region.
\end{description}

\begin{figure}[hbt!]
\begin{center}
\centerline{
\includegraphics[width=8cm,height=7.5cm]{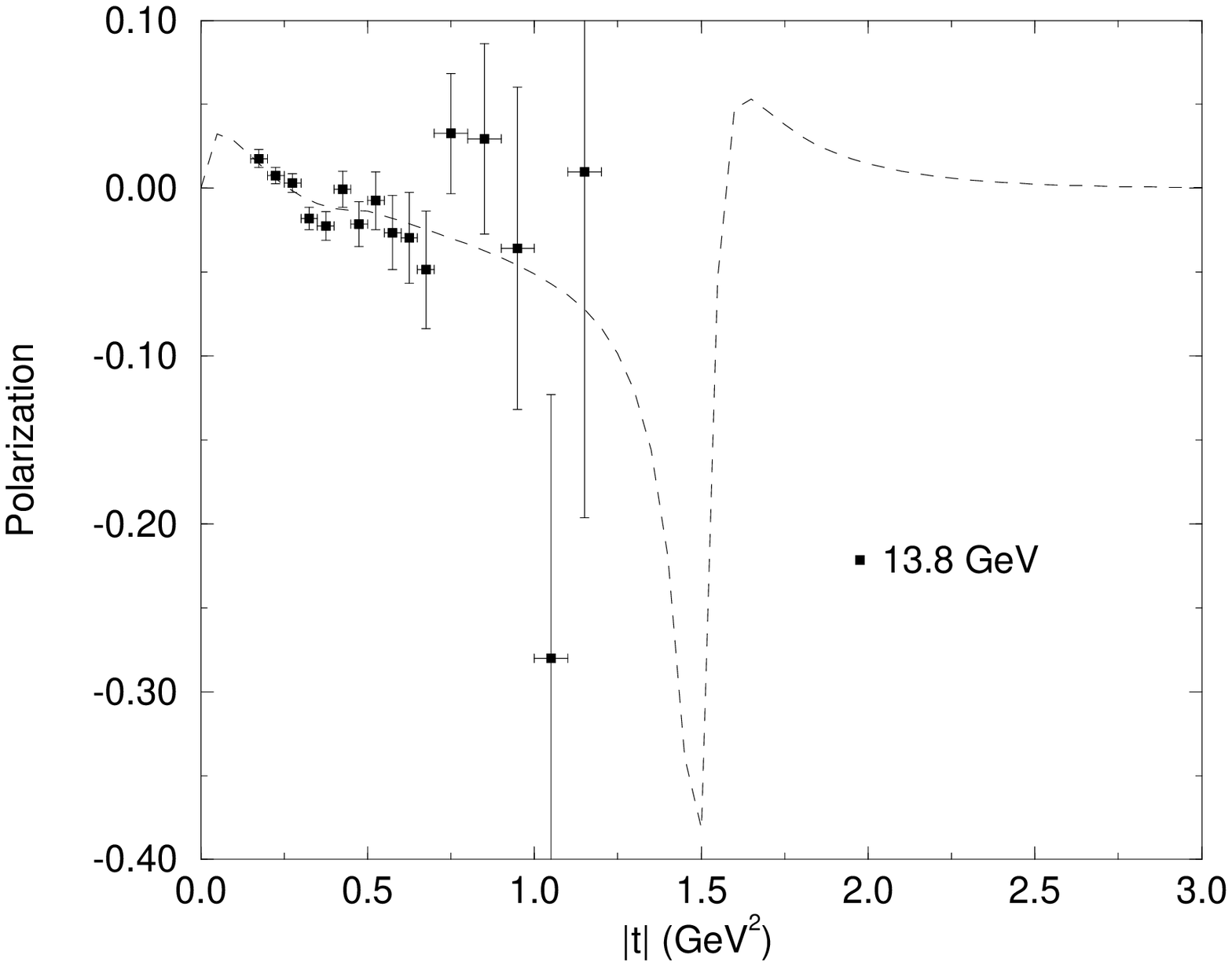}
\includegraphics[width=8cm,height=7.5cm]{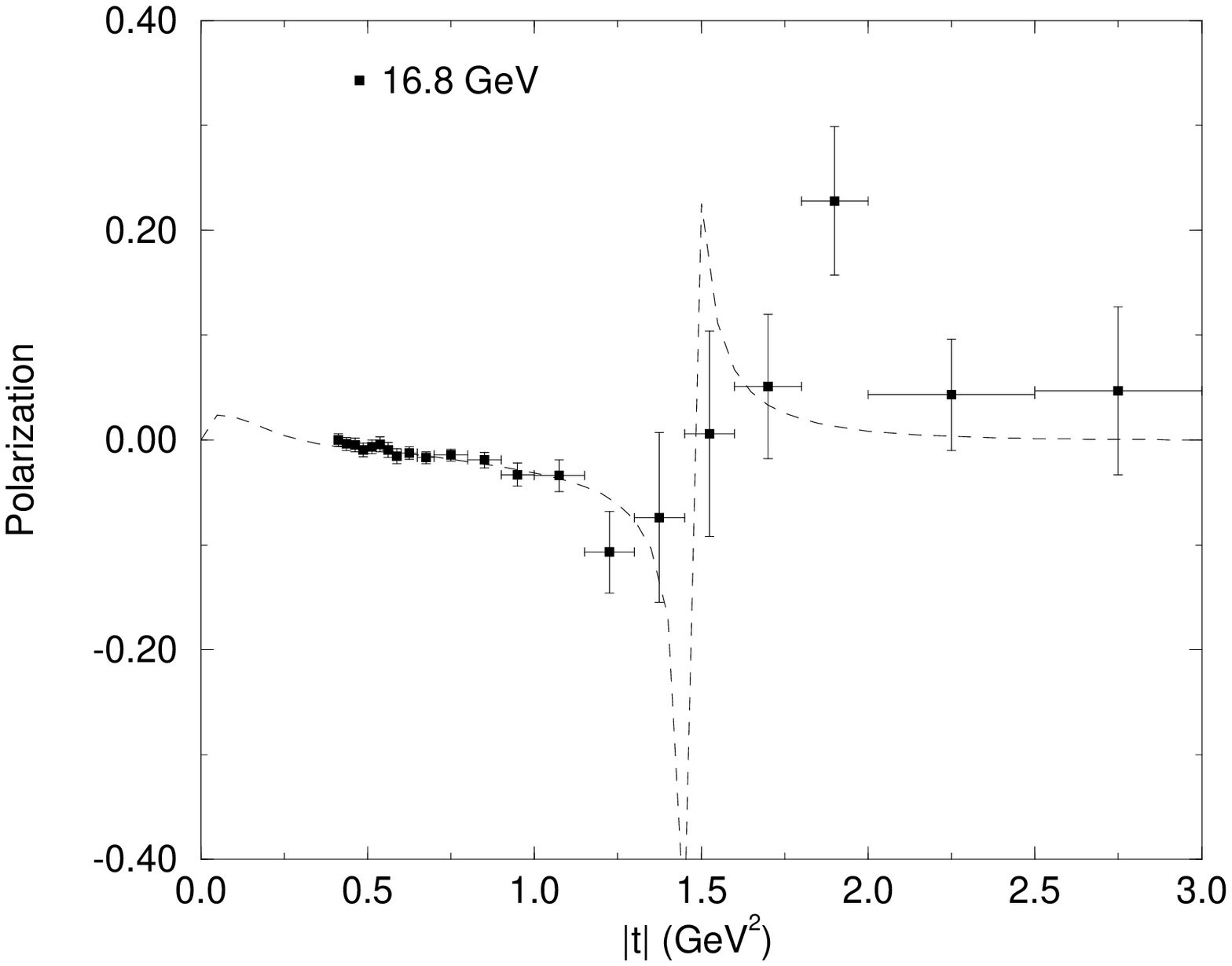}
}
\vspace{-0.5cm}
\centerline{
\includegraphics[width=7cm,height=7cm]{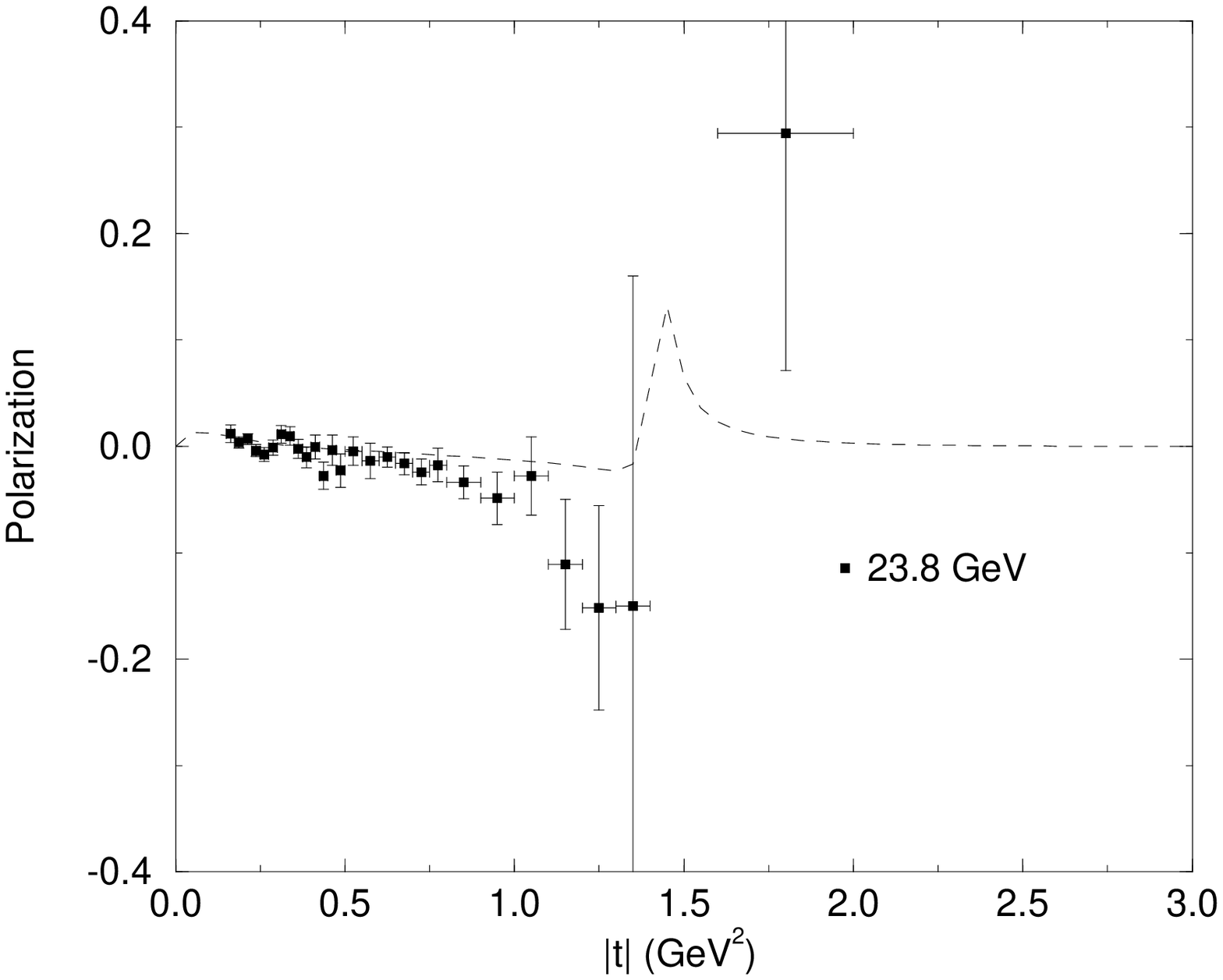}} 
\vspace{-0.5cm}
\caption{ Results from fitting polarization data 
at various energies (13.8, 16.8 and 23.8 GeV) without diffraction 
(parameters of Table \protect\ref{tab7param}).
}
\label{figpol7param}
\end{center}
\end{figure}

\begin{figure}[hbt!]
\begin{center}
\centerline{
\includegraphics[width=8cm,height=7.5cm]{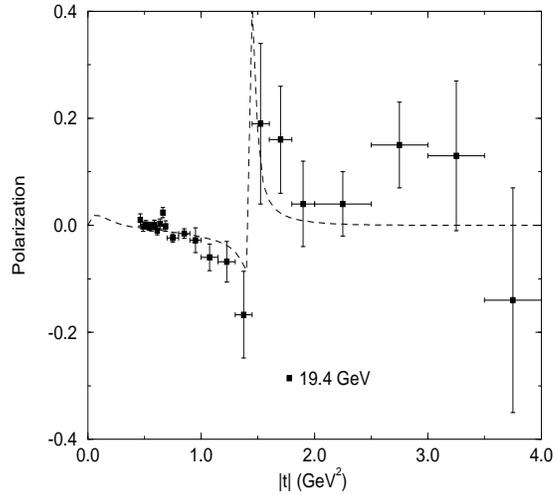}}
\caption{ The prediction for polarization at 19.4 GeV
compared with the experimental data using the parameters of 
Table \protect\ref{tab7param} (dashed line).
}
\label{figpol19pt4II}
\end{center}
\end{figure}

\begin{figure}[hbt!]
\begin{center}
\centerline{
\includegraphics[width=8cm,height=7.5cm]{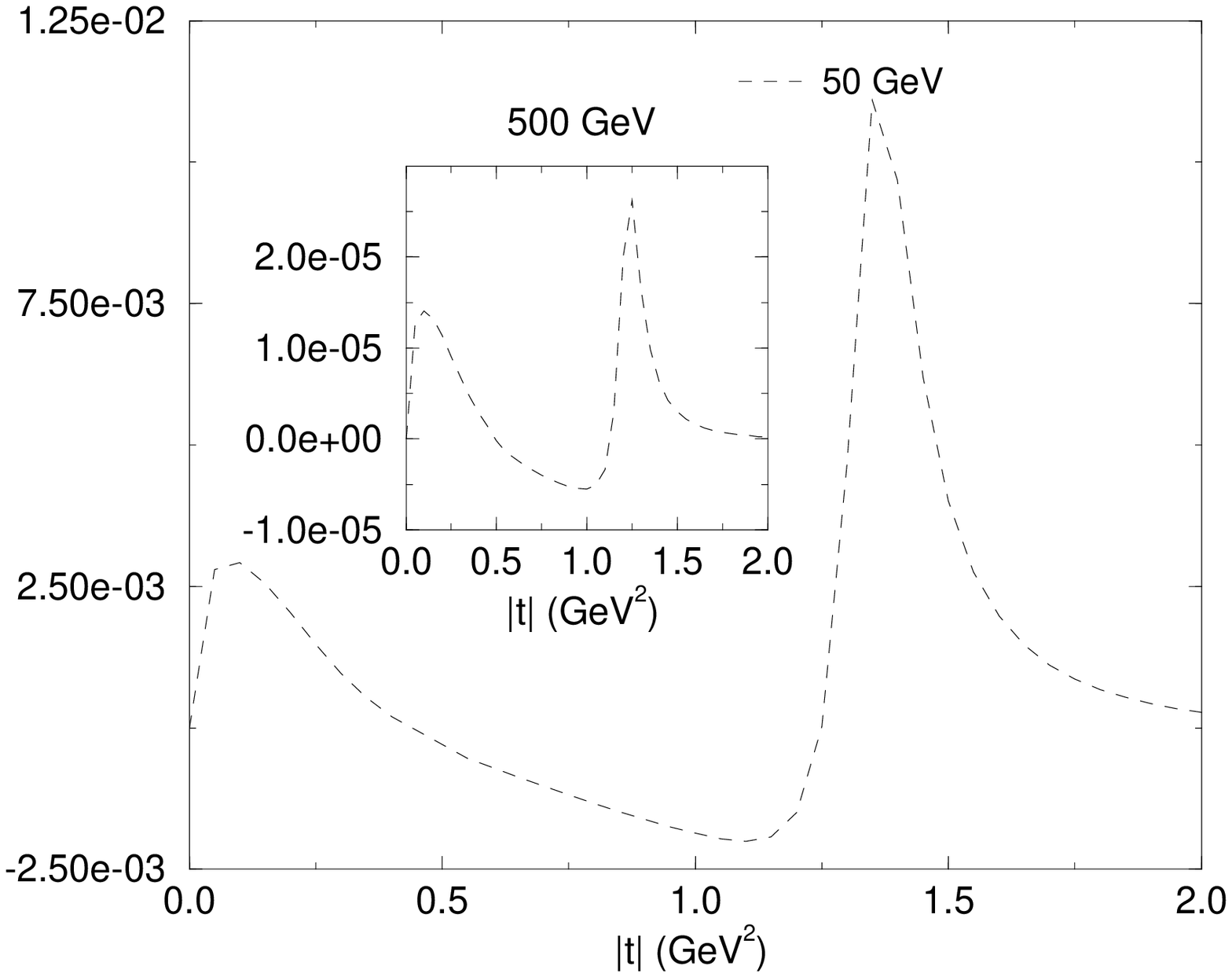}
}
\caption{The polarization predictions for 50 and 500 GeV
using the parameters of Table \protect\ref{tab7param} ({\it i.e.}, 
without diffraction). A detailed view of the 500 GeV is shown in the 
inset.
}
\label{figpol505007par}
\end{center}
\end{figure}

One open question remains the possibility that $\alpha^{sf}(0)$
be not the same as in the spin-non-flip amplitude. If $\alpha^{sf}(0)$
is left free to vary below unity ({\it i.e.}, no Pomeron contribution), 
one can still fit the data (albeit with a negative $\alpha^{sf}(0)$) 
and the description of the polarization remains essentially 
the same. The quality of the fit (the $\chi^2/d.f.$) is practically 
the same as can be seen in Fig. \ref{figpol7param} and 
\ref{figpol19pt4II}. Table \ref{tab7param} 
shows the values obtained for the parameters of eq. (\ref{spinampl2}) 
when $\alpha^{sf}(0)\neq\alpha_{\IP}(0)$. On the other hand, in this case 
all parameters change drastically. In particular, the couplings 
$\gamma_i$ and $\delta_i$ become totally absurd so that we incline 
to discard entirely this solution. However, this can be left to the 
experiment. In this case, in fact, when the fit to the polarization is 
extrapolated from  $\sqrt{s}=23.8$ GeV to RHIC energies 
(Fig. \ref{figpol505007par}), the polarization predictions are negligibly 
small (and very similar to those obtained with $\sin\theta$, 
see Fig. 7 on \cite{martini00}).

\begin{table}[hbt!]
\centerline{
\begin{tabular}{|c|c|c|c|}
\hline
\multicolumn{4}{|c|}{$\alpha_{\IP}^{sf}(0)=-0.129$} \\
\hline
$\gamma_1$ & -1031 & $\gamma_2$ & -46.8 \\
$\delta_1$ & 187 & $\delta_2$ & 3.67 \\
$\beta_1\;({\rm GeV}^{-2})$ & 7.84 & $\beta_2
\;({\rm GeV}^{-2})$ & 2.29 \\ \hline
\multicolumn{4}{|c|}{$\chi^2/d.f.=1.1$}\\
\hline
\end{tabular}
}
\caption{Values of the parameters in the absence of 
diffraction ($\protect\alpha^{sf}(0)<\alpha_{\IP}(0)$).
}
\label{tab7param}
\end{table}

\section{Conclusions}
\label{sec:concl}

We performed the fit of polarization data for $pp$ scattering 
with a Pomeron spin-flip amplitude where the kinematical zero is 
removed by the factor $\sqrt{-t}$, instead of $\sin\theta$ as 
done earlier \cite{martini00}. The motivation 
for this reanalysis is that, hidden inside the $\sin\theta$ option, 
there is a $1/\sqrt{s}$ dependence which affects the extrapolation to 
RHIC energies.

Although $\sqrt{-t}$ is more in line with a Regge parametrization, 
we cannot find a clear cut theoretical argument against using 
$\sin\theta$ in the 
Pomeron spin-flip amplitude but the differences predicted in the 
high energy extrapolations obtained here and in Ref. \cite{martini00} 
marks the importance of RHIC to solve this question.

If the kinematical zero is removed using $\sqrt{-t}$ and diffraction 
is present in the spin-flip amplitude, 
the Pomeron contribution to the spin-flip amplitude is 
smaller than to the spin-non-flip amplitude (about 10\%). 
At the same time, the extrapolation to RHIC energies is predicted 
to be non-negligible and considerably larger than using $\sin\theta$. 
The slope $\beta_1$ 
is somewhat smaller ($\beta_1=4.74\;{\rm GeV}^{-2}$ instead of 
6.25 ${\rm GeV}^{-2}$ in \cite{martini00}) but $\beta_2$ has 
approximately the same value ($\beta_2=2.29\;{\rm GeV}^{-2}$ instead of 
2.30 ${\rm GeV}^{-2}$ in \cite{martini00}). The $P$ values at 
$\sqrt{s}\sim 500$ GeV may be accessible to measurement in this case 
since they can reach 10 percent near the region of the dip in $d\sigma/dt$.

We conclude that, only in the case of a diffractive (Pomeron) contribution 
to the spin-flip $pp$ amplitude, will RHIC be able to obtain a measurable 
polarization. Such a contribution was suggested to be present more than 
30 years ago \cite{predazzi67} and is not ruled out by present data. On 
the other hand, the fit without diffractive contribution in the 
spin-flip amplitude leads to a solution whose set of parameters is 
quite absurd.

{\bf Acknowledgements.} 
One of us (AFM) would like to thank the Department of Theoretical Physics
of the University of Torino for its hospitality and the FAPESP of Brazil
for its financial support. Several discussions with Prof. E. Martynov and 
Prof. M. Giffon and the comments of Prof. O.V. Selyugin are gratefully
acknowledged.

\vspace{1.0cm}
\centerline{\bf APPENDIX}
\appendix
\section{The spin-non-flip amplitude}
\label{appe:spin}

The spin-non-flip amplitude utilized in this work is
\begin{equation}
a^{nf}(s,t)\equiv a_{+}(s,t)-a_{-}(s,t) ,
\end{equation}
where
\beq
a_{+}(s,t)=a_{\IP}(s,t)+a_f(s,t) 
\eeq
and
\beq
a_{-}(s,t)=a_{O}(s,t)+a_{\omega}(s,t).
\eeq

The expressions for the two Reggeons used in \cite{desgrolard00} are
\beq
a_R(s,t)=a_R\tilde{s}^{\alpha_R(t)}e^{b_R t},
\eeq
and
\beq
\alpha_R(t)=\alpha_R(0)+\alpha_R't,\; (R=f\; {\rm and} \; \omega)
\eeq
with $a_f (a_{\omega})$ real (imaginary). 

For the Pomeron, the non spin-flip amplitude is
\beq
a_{\IP}^{(D)}(s,t)=a_{\IP}\tilde{s}^{\alpha_{\IP}(t)}
[e^{b_{\IP}(\alpha_{\IP}(t)-1)}(b_{\IP}+ln\tilde{s})+d_{\IP}ln\tilde{s}] 
\eeq
while for the Odderon, we choose
\beqa
a_{O}(s,t)&=&(1-\exp(\gamma t))a_O\tilde{s}^{\alpha_O(t)}\nonumber\\
&\times &[e^{b_O(\alpha_O(t)-1)}(b_O+ln\tilde{s})+d_Oln\tilde{s}],
\eeqa
and again $a_{\IP} (a_O)$ real (imaginary). We use 
$\alpha_i(t)=\alpha_i(0)+\alpha_i't$ where $i=\IP ,O$.

Our definition for the amplitude follows \cite{desgrolard00} so 
that
\beqa
\sigma_t={4\pi\over s}{\rm Im}\{ a^{nf}(s,t=0)\},
\label{sctot} \\
{d\sigma\over dt}={\pi\over s^2}(|a^{nf}(s,t)|^2+2|a^{sf}(s,t)|^2).
\label{scdif}
\eeqa

In this work we retain the same parameters for the spin-non-flip
amplitude as in \cite{desgrolard00} and we keep them fixed while 
fitting the parameters of the spin-flip 
amplitude. We utilize the dipole model at the Born level since 
most of the polarization data is contained in the $t$-domain 
corresponding to the region before the dip in $d\sigma/dt$ (well 
described without eikonalization). The values of the parameters of 
the spin-non-flip amplitude \cite{desgrolard00} are shown in 
Table \ref{tabsnfpar}.

\begin{table}[hbt!]
\centerline{
\begin{tabular}{|c|c|c|c|c|}
\hline
 & Pomeron & Odderon & $f$-Reggeon & $\omega$-Reggeon \\ 
\hline
$\alpha_i(0)$ & 1.071 & 1.0 & 0.72 & 0.46 \\
$\alpha_i'$ & 0.28 ${\rm GeV}^{-2}$ & 0.12 ${\rm GeV}^{-2}$ & 
0.50 ${\rm GeV}^{-2}$ & 0.50 ${\rm GeV}^{-2}$ \\
$a_i$ & -0.066 & 0.100 & -14.0 & 9.0 \\
$b_i$ & 14.56 & 28.10 & 1.64 ${\rm GeV}^{-2}$ & 0.38 ${\rm GeV}^{-2}$ \\
$d_i$ & 0.07 & -0.06 & - & - \\ 
$\gamma$ & - & 1.56 ${\rm GeV}^{-2}$ & - & - \\
\hline
\end{tabular}
}
\caption{Parameters of the dipole model at the Born level 
with $i=\IP ,O,f,\omega$ (from Ref. \protect\cite{desgrolard00}).
}
\label{tabsnfpar}
\end{table} 

To calculate the polarization we utilized the form
\beq
P=2
{{\rm Im}(a^{nf}(s,t)(a^{sf}(s,t))^{\star})\over 
|a^{nf}(s,t)|^2+2|a^{sf}(s,t)|^2} ;
\eeq
where the star on the numerator means complex conjugate.

\end{document}